\documentclass[11pt]{article}
\usepackage{graphicx}
\usepackage{setspace}
\begin{document}
\bibliographystyle{unsrt}
\def\ra{\rangle}
\def\la{\langle}
\def\aao{\hat{a}}
\def\aaot{\hat{a}^2}
\def\aco{\hat{a}^\dagger}
\def\acot{\hat{a}^{\dagger 2}}
\def\ano{\aco\aao}
\def\bao{\hat{b}}
\def\baot{\hat{b}^2}
\def\bco{\hat{b}^\dagger}
\def\bcot{\hat{b}^{\dagger 2}}
\def\bno{\bco\bao}
\def\beqn{\begin{equation}}
\def\eeqn{\end{equation}}
\def\bear{\begin{eqnarray}}
\def\eear{\end{eqnarray}}
\def\cdott{\cdot\cdot\cdot}
\def\bcen{\begin{center}}
\def\ecen{\end{center}}
\def\nbar{\bar{n}}
\def\eps{\epsilon}
\def\hrho{\hat{\rho}}
\def\rhom{\hat{\rho}_m}
\def\rhot{\hat{\rho}_t}
\def\rhod{\hat{\rho}_d}
\title{Truncated coherent states and photon-addition}
\author{
S. Sivakumar\\Materials Physics Division\\ 
Indira Gandhi Centre for Atomic Research\\ Kalpakkam 603 102 INDIA\\
Email: siva@igcar.gov.in\\}
\maketitle
\begin{abstract}
A class of  states of the electromagnetic field involving superpositions of all the excited states above a specified low energy eigenstate of the electromagnetic field is introduced.  These states and the photon-added coherent states are shown to be the limiting cases of a generalized photon-added coherent states.  This new class of states is nonclassical, non-Gaussian and has equal uncertainties in the field quadratures. For suitable choices of parameters, these uncertainties are very close to those of the coherent states.  Nevertheless, these states exhibit sub-Poissonian photon number distribution, which is a nonclassical feature.   Under suitable approximations, these states become the generalized Bernoulli states of the field.  Nonclassicality of these states is quantified using their entanglement potential.    

\end{abstract}
\newpage
\section{Introduction}

    Quantum theory stipulates a lower bound on the  product of the uncertainties of in a pair of non-commuting observables such as the position and momentum, the phase and particle number, etc.  In the context of the position-momentum uncertainty relation $\Delta x\Delta p\ge \hbar/2$, there are states which satisfy $\Delta x\Delta p = \hbar/2$. Such states are called minimum uncertainty states(MUS), for example, the coherent states and the squeezed vacuum. Of all the MUS,  the coherent states $\vert\alpha\ra$\cite{glauber2,ecg},
\beqn
\vert\alpha\ra=\exp\left(-\frac{\vert\alpha\vert^2}{2}\right)\sum_{n=0}^\infty \frac{\alpha^n}{\sqrt{n!}}\vert n\ra,
\eeqn
defined for every complex number $\alpha$,  
are very special as their phase space distributions (P- and Wigner-distributions) are well-defined probability densities\cite{gerryknight}.  For that reason, the coherent states are considered to be "classical" among the quantum states.  Any other pure state of a harmonic oscillator is non-classical, in the sense that its phase space distribution (P-function or Wigner) fails to be non-negative.  Interestingly, superpositions of two coherent states  generate non-classical states; the even and odd coherent states\cite{yurkestoler, eocs, ourjou} are prime examples of such states. In the case of the electromagnetic field, photon-addition and state truncation  are two other processes  to  create nonclassical states from coherent states\cite{gsatara, nha}.   Both these routes have been experimentally realized\cite{zavatta, babichev, grangier, marek}. In the process of state truncation, a finite number of Fock states are retained in the coherent states.  Truncated coherent states (TCS) are defined as\cite{kuang, leonski, tanas94, leonski97}
\beqn
\vert\alpha,N;u\ra=N_u\sum_{n=0}^N\frac{\alpha^n}{\sqrt{n!}}\vert n\ra,
\eeqn
where $N_u^{-2}$ is $\exp(\vert\alpha\vert^2)[1-\gamma(N+1,\vert\alpha\vert^2)/N!]$.   The incomplete Gamma function $\gamma(N,x)$ \cite{grad}is defined as 
\beqn\label{ingamma}
\gamma(N,x)=(N-1)!\left[1-\exp(-x)\sum_{j=0}^{N-1}\frac{x^j}{j!}\right],
\eeqn
$N$ being the order of the function.  Since there is an upper limit on the number of Fock states involved in the superposition, this state is referred as upper-truncated coherent states (UTCS).  
These experimentally realizable states have been studied extensively for their nonclassical features\cite{leonski, ppb, kuang_pla, jcp1, jcp2}.  In the limit $N\rightarrow\infty$, $\vert\alpha,N;u\ra$ becomes the coherent state $\vert\alpha\ra$.  Another class of states associated with the UTCS is defined as follows:
\beqn\label{Ltcsdef}
\vert\alpha,N;l\ra=N_l\sum_{n=0}^\infty\frac{\alpha^n}{\sqrt{(N+1+n)!}}\vert N+1+n\ra.
\eeqn
The normalization constant satisfies $N_l^{-2}=\exp(\vert\alpha\vert^2)\gamma(N+1,\vert\alpha\vert^2)/N!$.
The lower limit on the Fock states implies that these states could be aptly termed  as lower-truncated coherent states (LTCS).  By construction, the UTCS and LTCS are orthogonal to each other.  If $\alpha\rightarrow 0$, then $\vert\psi,N\ra_u$ and $\vert\psi,N\ra_l$ are $\vert 0\ra$ and $\vert N+1\ra$ respectively.
The canonical coherent state $\vert\alpha\ra$ is expressed as 
\beqn
\vert\alpha\ra=\frac{\exp(-\vert\alpha\vert^2/2)}{N_uN_l}\left[N_l\vert\alpha,N;u\ra+N_u\alpha^{N+1}\vert\alpha,N;l\ra\right].
\eeqn

In this work, the properties of the LTCS are presented and compared with those of the UTCS.  In Section II, the relation between the LTCS and another well known nonclassical state, namely, the photon-added coherent states (PACS) is established.  Nonclassical features such as squeezing and sub-Poissonian statistics are discussed in Section III, followed by a summary of the results in Section IV.

\section{LTCS and Photon-added coherent states}

  Photon-addition is mathematically represented by the action of a suitable creation operator on a state of the electromagnetic field.  The process leads to generation of nonclassical states from the coherent states\cite{gsatara}.   The process and its various generalizations have been studied both experimentally and theoretically\cite{zavatta}; see \cite{kim_pacs} for a recent review.  

 Photon-added coherent states (PACS) are defined as\cite{gsatara}
\beqn\label{pacs}
\vert\alpha,m\ra\propto\aao^{\dagger m}\vert\alpha\ra=\exp(-\frac{\vert\alpha\vert^2}{2})\sum_{n=0}^\infty\frac{\alpha^n\sqrt{(n+m)!}}{n!}\vert n+m\ra.  
\eeqn
The parameter $m$ is the order of the PACS.  These states have been experimentally realized in optical parametric down-conversion\cite{zavatta}. 
Both the LTCS $\vert\alpha,N'l\ra$ and the PACS $\vert\alpha,m\ra$ with $m=(N+1)$ are defined  are superpositions of the the same set of Fock states $\{\vert n\ra\}$ $[n=m+1,~m+2,\cdots]$.  To see the connection between the two states, consider the  deformed annihilation operator $\hat{A}=(1+k\ano)\aao$, where $k$ is nonnegative, not exceeding unity.  Associated with $\hat{A}$ is the "deformed creation operator" $\hat{B}^\dagger$ $=\aco(1+k\ano)^{-1}$, and these two operators satisfy  $\left[\hat{A},\hat{B}^\dagger\right]=I$\cite{shanta}.  It is important to note that  $\hat{B}^\dagger$ is not the adjoint of $\hat{A}$.  Starting with the coherent state  $\vert\alpha\ra$, define
\beqn
\vert\alpha,m\ra_k=N_k\hat{B}^{\dagger m}\vert\alpha\ra=N_k\sum_{n=0}^\infty\frac{\alpha^n}{n!}\frac{\sqrt{(n+m)!}}{\Pi_{j=0}^{m-1}(1+kn+kj)}\vert n+m\ra.
\eeqn
In the limit of vanishing $k$, the state $\vert\alpha,m\ra_k$ becomes the $m$-PACS $\vert\alpha,m\ra$; with $k=1$, the state becomes the LTCS.  Hence, the PACS and LTCS are the limiting cases of the deformed-PACS $\vert\alpha,m\ra_k$.  If $k=1$, the creation operation $\hat{B}^\dagger=\aco(1+\ano)^{-1}$ is the right inverse of the annihilation operator $\aao$, that is, $\aao\hat{B}^\dagger=I$\cite{roymehta}.   Thus, the LTCS are the obtained by the repeated application of the right inverse of $\aao$ on the coherent state $\vert\alpha\ra$.   The notion of deformed photon-addition has been discussed in the context of Posch-Teller potential\cite{dpacs} too.
The overlap between the  PACS and LTCS of same amplitude $\alpha$ and order $N+1$ is
\beqn
\vert\la\alpha,N+1\vert\alpha,N;l\ra_l\vert^2=\frac{\vert\alpha\vert^{2(N+1)}}{(N+1)\gamma(N+1,\vert\alpha\vert^2)L_{N+1}(-\vert\alpha\vert^2)},
\eeqn
where $L_N(x)$ is the Laguerre polynomial of order $N$\cite{grad}.  
For large values of $N$, the overlap becomes negligible indicating that the two limiting cases are nearly orthogonal to each other.

      It is known that the PACS of order $N+1$ provide a resolution of identity for the subspace spanned by the number states ${\vert N+1\ra,~ \vert N+2\ra,~\vert N+3\ra, \cdots}$\cite{penson}.  A similar relation holds for the LTCS too.  In this case,
\beqn
\frac{1}{\pi}\int d^2\alpha \frac{\gamma(N+1,\vert\alpha\vert^2)}{N!}\vert\alpha,N;l\ra\la\alpha,N;l\vert =I-\sum_{n=0}^N\vert n\ra\la n\vert.
\eeqn
The RHS of the above equation is the identity operator on the subspace spanned by $ \vert N+1\ra,\vert N+2\ra,{\vert N+3\ra,\cdots}$.  From the definition of the incomplete Gamma function given in Eq. \ref{ingamma}, it is seen that $\gamma(N+1, x)$ is non-negative if $x \ge 0$.  This means that the weight function in the resolution of identity by the LTCS is nonnegative. 

     The result that a LTCS of order $N+1$ is obtained by the action of the deformed creation operator $\hat{B}^{\dagger N}$ on the coherent state $\vert\alpha\ra$ leads to a plausible method of generating these states, in close parallel to the suggestion by Agarwal and Tara\cite{gsatara} to generate the PACS.  Consider a two-level atom interacting with a single-mode of the electromagnetic field.  The Hamiltonian in the interaction picture is taken to be $H_I=\hbar\lambda\left[\hat{B}^\dagger\sigma_-+\hat{B}\sigma_+\right]$, which is a generalized  Jaynes-Cummings model of atom-field interaction\cite{buck, welsch}. This  interaction corresponds to an intensity-dependent atom-field coupling.  Here $\sigma_+$ and $\sigma_-$ are respectively the raising and lowering operators for the atomic states $\vert e\ra$ and $\vert g\ra$ and  $\lambda$ is coupling constant.  The actions of the raising and lowering operators on the atomic states are given by $\sigma_\pm\vert e\ra=(\vert g\ra\mp\vert g\ra)/2$ and $\sigma_\pm\vert g\ra=(\vert e\ra\pm\vert e\ra)/2$.   The initial state of the atom-field state is $\vert\alpha\ra\vert e\ra$; the atom in the excited state $\vert e\ra$ and the field in the coherent state $\vert\alpha\ra$.  If the interaction duration is small ($\vert t\lambda\vert<<1$), then the evolution operator is approximated to 
\beqn
\exp(-iH_It/\hbar)\approx I-it\lambda\left[\hat{B}^\dagger\sigma_-+\hat{B}\sigma_+\right].
\eeqn
The time-evolved state of the atom-field system is the superposition $\vert\alpha\ra\vert e\ra-it \lambda\hat{B}^\dagger\vert\alpha\ra\vert g\ra$, an entangled state of the system.  Subsequent to the interaction, if the atom is detected in its ground state $\vert g\ra$, the field is in the state $\hat{B}^\dagger\vert\alpha\ra$, which is the LTCS with $N=0$.  If another two-level atom in its excited state   interacts with this resultant field $\hat{B}^\dagger\vert\alpha\ra$, and the the atom is detected in its ground state after the interaction, the field changes to $\hat{B}^{\dagger 2}\vert\alpha\ra$.  After interacting with a sequence of $N$ atoms which are subsequently detected in their respective ground states, the initial coherent state of the field is transformed to $\hat{B}^{\dagger N}\vert\alpha\ra$, the LTCS of order $N$.  It is possible to make the atom-field coupling to be time-dependent  and its form can be tailored to generate arbitrary superpositions\cite{eberly}.     Another route to realize the type of the intensity-dependent coupling is to consider situations in which  the rotating-wave approximation is not suitable and the resulting Hamiltonian is equivalent to including an intensity-dependent interaction of the form $(1-k\ano)\aao$, which approximates  the operator $(1+k\ano)^{-1}\aao$ when $k\la\ano\ra<1$\cite{naderi}, wherein the expectation value $\la\ano\ra$ is in the state $\vert\alpha,N;l\ra$.   Further, it has been  shown that the properties of a cavity containing nonlinear media   can be tuned to provide different forms of intensity-dependent atom-field couplings\cite{teppo}.

    Another state that is relevant in the context of the commutation $\left[\hat{A},\hat{B}^\dagger\right]=I$ is $\exp(\alpha\hat{B}^\dagger-\alpha^*\hat{A})\vert 0\ra$, whose Fock basis expansion is 
\beqn
\exp(\alpha\hat{B}^\dagger-\alpha\hat{A})\vert 0\ra\propto \sum_{n=0}^\infty\frac{\alpha^n}{\sqrt{n!}}\frac{1}{(1+k)(1+2k)\cdots(1+(n-1)k)}\vert n\ra.
\eeqn 
Since $\hat{B}^\dagger$ is not the adjoint of $A$, the operator $\exp(\alpha\hat{B}^\dagger-\alpha^*\hat{A})$ is not unitary. 
However,  its action on the vacuum results in a state that is normalizable for all values of $\alpha$.   In this work, these states are not discussed further.
\section{Nonclassical features}

Pure states of the electromagnetic field, except the coherent states, are nonclassical.    
Squeezing and sub-Poissonian statistics are two of the experimentally verifiable nonclassical features that a state of the electromagnetic field can exhibit.  In this section, these two aspects of the LTCS are discussed.  
If the deforming parameter $k=1$,  the expectation of the annihilation operator and creations operators are
\bear
\la\aao\ra_u&=&\la\aco\ra^*_u=N\alpha \frac{(N-1)!-\gamma(N-1,\vert\alpha\vert^2)}{N!-\gamma(N,\vert\alpha\vert^2)}\\
\la\aao^2\ra_u&=&\la\aao^{\dagger 2}\ra^*_u=N(N-1)\alpha^2 \frac{(N-1)!-\gamma(N-1,\vert\alpha\vert^2)}{N!-\gamma(N,\vert\alpha\vert^2)},\\
\la\aao\ra_l&=&\la\aco\ra^*_l=\alpha,\\
\la{\aao}^2\ra_l&=&\la\aao^{\dagger 2}\ra^*_l=\alpha^2.\
\eear
The subscripts $u$ and $l$ refer to the UTCS and LTCS respectively.  
An interesting feature is  that the expectations values of the two operators in LTCS are independent of $N$.  However, the expectation value of the number operator $\ano$ does not possess this feature,
\bear
\la\ano\ra_u&=&N\frac{(N-1)!-\gamma(N-1,\vert\alpha\vert^2)}{N!-\gamma(N,\vert\alpha\vert^2)}\vert\alpha\vert^2\\
\la\ano\ra_l&=&N\vert\alpha\vert^2\frac{\gamma(N,\vert\alpha\vert^2)}{\gamma(N+1,\vert\alpha\vert^2)}
\eear  

\subsection{Squeezing in X and P quadratures}

\begin{figure}
\centering
\includegraphics[height=8cm,width=9cm]{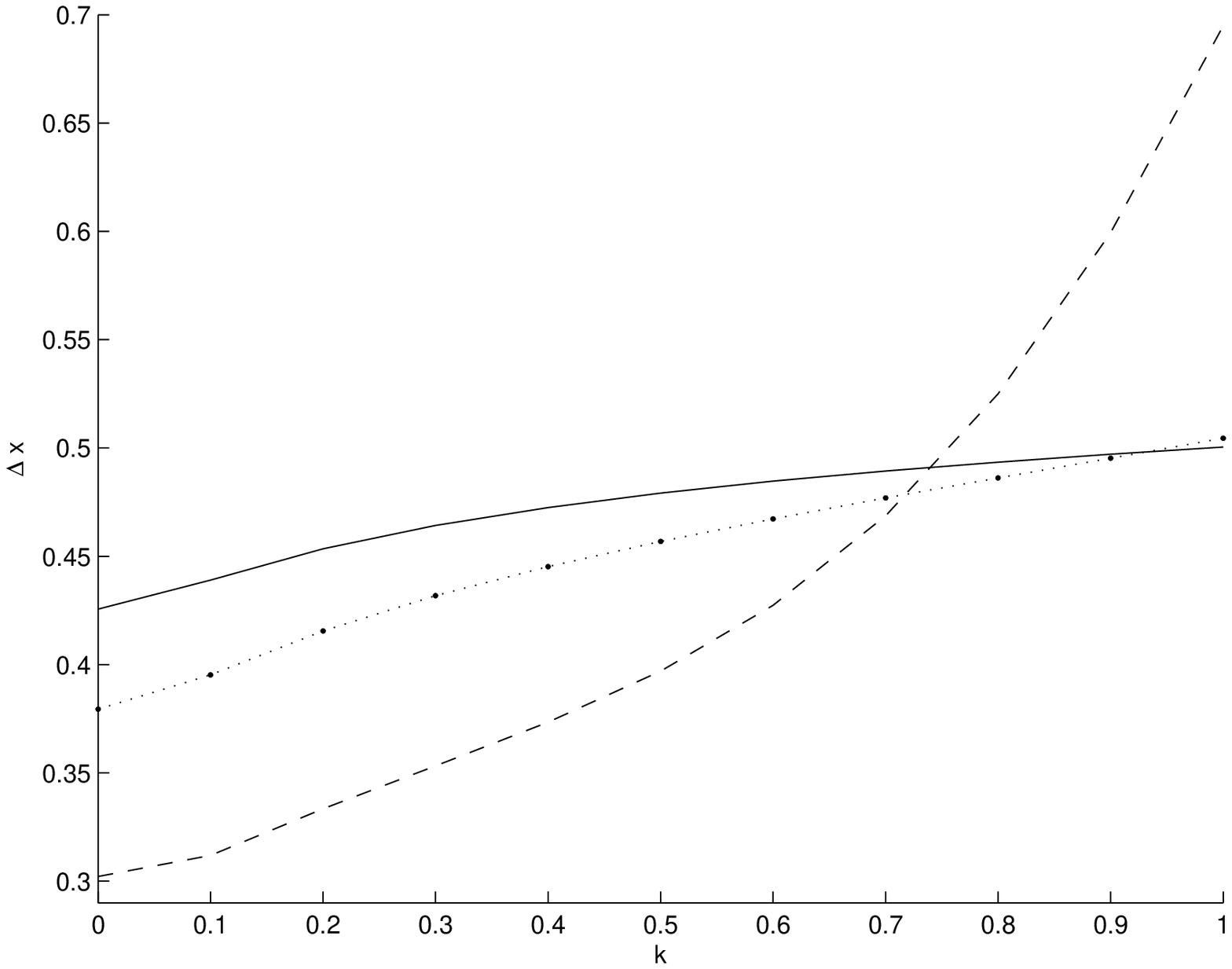}
\caption {Uncertainty in $X$-quadrature as a function of the parameter $k$ for states $\vert\alpha, N\ra_k$, with $N=2$ (continuous), $N=3$ (line with dots) and $N=5$ (dashed). In all cases $\alpha=\sqrt{10}$.  The limiting cases corresponding to $k=0$ and 1 are the PACS and LTCS respectively. }
\label{fig:FigI}
\end{figure}

For the single mode electromagnetic field, the following quadratures are defined,
\beqn
\hat{X}=\frac{\aco+\aao}{\sqrt{2}}~~~~\hat{P}=\frac{\aao-\aco}{i\sqrt{2}}.
\eeqn
The corresponding uncertainties are $(\Delta X)^2=\la\hat{X}^2\ra-\la\hat{X}\ra^2$ and $(\Delta P)^2=\la\hat{P}^2\ra-\la\hat{P}\ra^2$.  For the coherent states of the electromagnetic field, 
uncertainties  in the two quadratures are  equal to 1/2. 
A state is said to exhibit squeezing in a quadrature if the uncertainty  in that quadrature is smaller than 1/2.  The PACS, which corresponds to $k=0$ in $\vert\alpha,m\ra_k$, exhibits squeezing.  
In Fig. \ref{fig:FigI}, the uncertainty in $X$ quadrature is shown as a function of the parameter $k$, for three different values of $N$.     The lowest values of $\Delta X$ are achieved for the PACS ($k=0$ case in the figure).  However, squeezing occurs over a substantial range of $k$.  If $N < \vert\alpha\vert^2$, the amount of squeezing in the $X$-quadrature is larger for smaller $N$. The range of $k$ over which squeezing occurs, however, decreases as $N$ increases.    

\begin{figure}
\centering
\includegraphics[height=8cm,width=10cm]{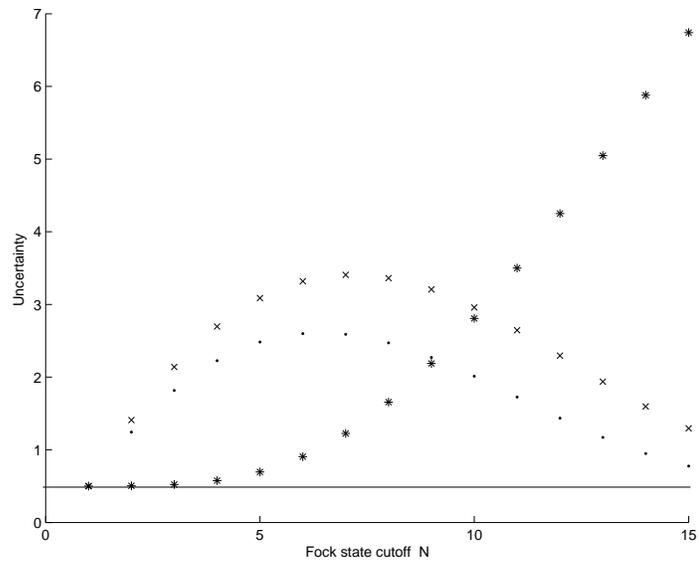}
\caption {Uncertainty profiles of $X-$ and $P-$quadratures for the UTCS and LTCS as a function of the cutoff number $N$.  For the UTCS, $\Delta X$ ($\times$) and  $\Delta P$ (.)are shown; for LTCS $\Delta X$(*) is shown, which is also the profile for $\Delta P$.  In all cases $\alpha=\sqrt{10}$.}
\label{fig:FigIII}
\end{figure}

In this work, the focus is on the properties of the LTCS.  Explicit expressions for the uncertainties in the  LTCS are
\beqn
(\Delta X)^2_l=(\Delta P)^2_l=\frac{1}{2}\left[1+2\vert\alpha\vert^2\left[(N+1)\frac{\gamma(N,\vert\alpha\vert^2)}{\gamma(N+1,\vert\alpha\vert^2)}-1\right]\right].
\eeqn
In Fig. \ref{fig:FigIII} , the variation of $\Delta X$ and $\Delta P$ are shown for different values of $N$ for the LTCS corresponding to $\alpha=\sqrt{10}$. 

It is seen from the figures that the quantities $\Delta X$ and $\Delta P$ are nearly equal to those of the coherent state $\alpha$, for $N\le \vert\alpha\vert^2$, though the average number of photons is high.   If the condition on $N$ is satisfied, the states $\vert\psi\ra_l$ are MUS with equal variances in the quadratures, whereas the squeezed vacuum is a MUS though with unequal distribution of fluctuations.
Unlike in the case of the PACS,  the uncertainties in the two quadratures are equal for the LTCS, which holds for  the number states and coherent states too.  This, in turn, means that  LTCS cannot exhibit reduced fluctuation in any of the quadratures, whereas PACS exhibits squeezing.   \\  

For comparison, uncertainty profiles for the UTCS are also given in Fig. \ref{fig:FigIII}.  
In the state $\vert\psi,N\ra_u$,  squeezing occurs if the magnitude of $\alpha$ is small.  In that limit, UTCS  is well approximated by  a superposition of the vacuum state $\vert 0\ra$ and the single photon state $\vert 1\ra$.  Such states are known as Bernoulli states and they exhibit squeezing\cite{bernoulli}.  

The overlap $\vert\la\alpha\vert\alpha,N\ra_l\vert^2$ between the coherent state $\vert\alpha\ra$ and the UTCS is 
$\exp(-\vert\alpha\vert^2)\sum_{n=0}^N\vert\alpha\vert^{2n}/n!$. 
If $N >> \vert\alpha\vert^2$, the summation is nearly $\exp(\vert\alpha\vert^2)$ and the overlap is almost equal to unity.  In that case, the UTCS is almost  the coherent state $\vert\alpha\ra$, which is a MUS.  It is interesting to note that for the UTCS the maximum uncertainties in the quadratures occur when  $N\approx\vert\alpha\vert^2/\sqrt{2}$.

\subsection{Phase distribution}

\begin{figure}
\centering
\includegraphics[height=8cm,width=10cm]{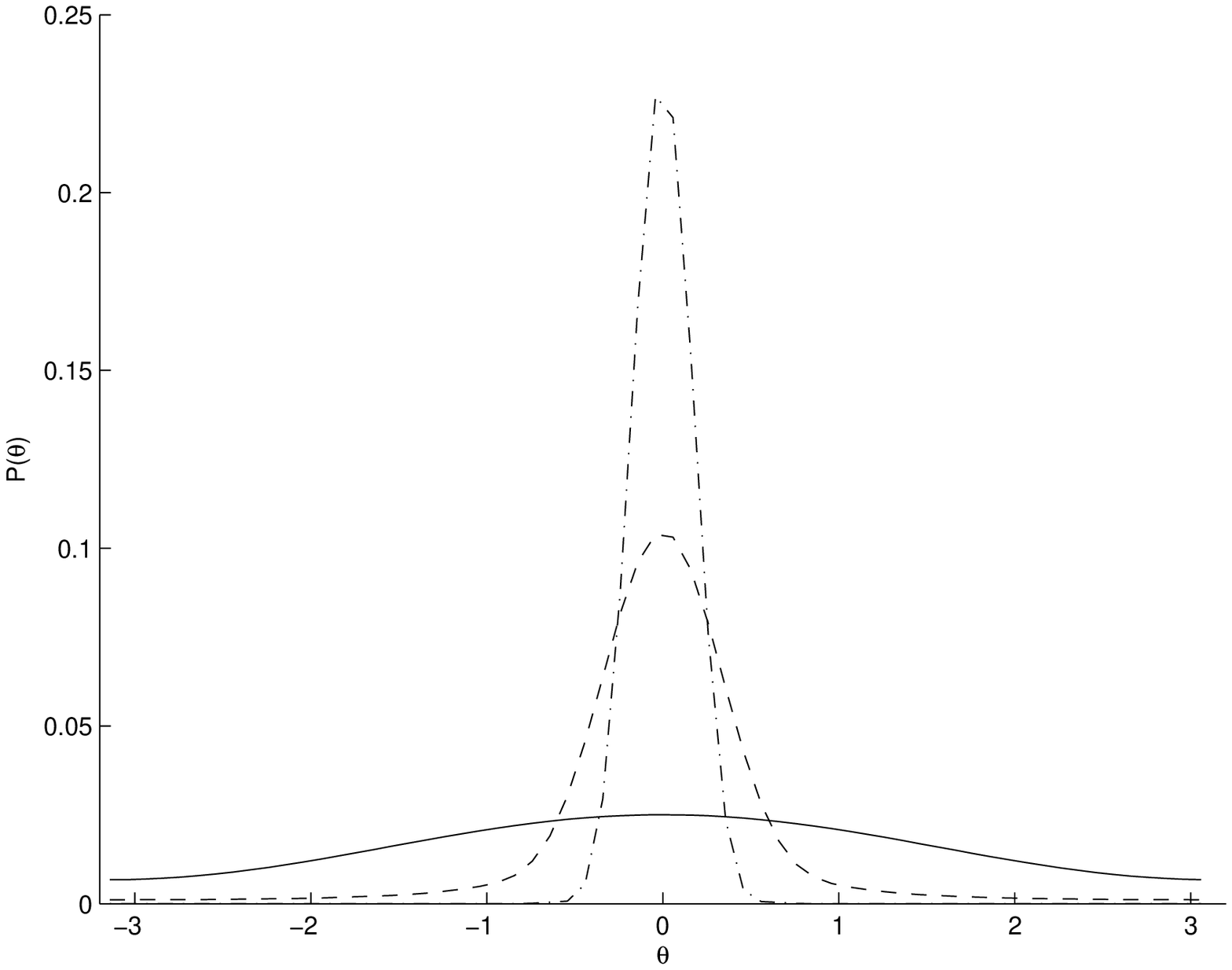}
\caption {Phase distribution $P_u(\theta)$ for the UTCS for two different values of the cutoff parameter, $N=2$ (continuous), $N=10$ (dashed) and $N=20$ (dot-dash).  The amplitude $\alpha=\sqrt{10}$.}
\label{fig:FigIV}
\end{figure}

\begin{figure}
\centering
\includegraphics[height=8cm,width=10cm]{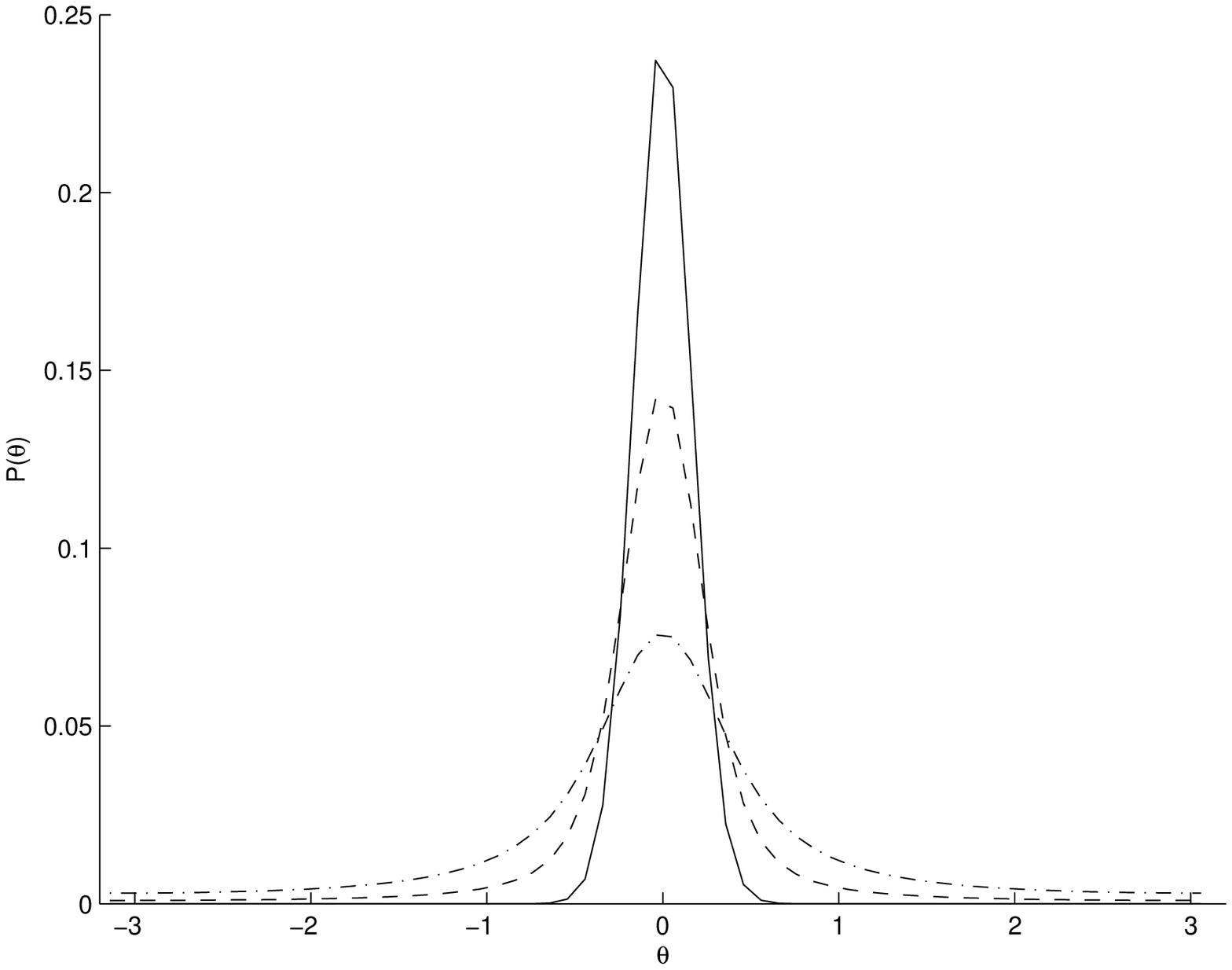}
\caption {Phase distribution $P_l(\theta)$ for the LTCS for two different values of the cutoff parameter, $N=2$ (continuous), $N=10$ (dashed) and $N=20$(dot-dash). The amplitude $\alpha=\sqrt{10}$.}
\label{fig:FigV}
\end{figure}

        Phase distribution of a quantum state of the electromagnetic field  provides information about uncertainty in the phase of the field.  This, in turn, imposes a lower limit on the the fluctuations in the number of photons.   An useful approach to define phase distribution is the Pegg-Barnett formulation \cite{qphase}.  Phase distribution of an arbitrary state of the field is obtained by the overlap of the state with the phase state 
\beqn
\vert\theta\ra=\frac{1}{\sqrt{1+s}}\sum_{n=0}^s e^{in\theta}\vert n\ra.
\eeqn
and taking the limit $s\rightarrow\infty$.
The respective phase distributions for the states $\vert\psi\ra_u$ and $\vert\psi\ra_l$ are
\beqn
P_u(\theta)=\lim_{s\rightarrow\infty}\frac{s+1}{2\pi}\vert\la\theta\vert\psi\ra_u\vert^2=N_u^2\left|\sum_{n=0}^N\frac{\exp(-in\theta)\alpha^n}{\sqrt{n!}}\right|^2,
\eeqn
and
\beqn
P_l(\theta)=\lim_{s\rightarrow\infty}\frac{s+1}{2\pi}\vert\la\theta\vert\psi\ra_l\vert^2=N_l^2\left|\sum_{n=N+1}^\infty\frac{\exp(-in\theta)\alpha^n}{\sqrt{n!}}\right|^2.
\eeqn
The phase distributions for the UTCS and LTCS are shown in  Figs. \ref{fig:FigIV} and \ref{fig:FigV} respectively, for three different values of $N$.  As the cutoff $N$ increases, the phase-distribution of UTCS  approaches that of a coherent state of amplitude $\alpha$.  In the case of LTCS,   
the phase-distribution broadens as $N$ increases.  This feature is shown in Fig. \ref{fig:FigV}.  If $N>>\vert\alpha\vert^2$, the LTCS is  a  superposition of the states  $\vert N+1\ra$ and $\vert N+2\ra$,
\beqn\label{bernoulli}
\vert\alpha,N;l\ra\approx\sqrt{\frac{N+2}{2+N+\vert\alpha\vert^2}}\left[\vert N+1\ra+\frac{\alpha}{\sqrt{N+2}}\vert N+2\ra\right].
\eeqn  
Such a superposition has a fairly well defined number of quanta, the corresponding  uncertainty in the number of photons is
\beqn
(\Delta n)^2=\la\ano\ano\ra-\la\ano\ra^2=\frac{\vert\alpha\vert^2(2+N)}{(2+\vert\alpha\vert^2+N)^2}.
\eeqn
If $N >> \vert\alpha\vert^2$, then $\Delta n<< 1$, that is, the photon number fluctuation is small.  Consequently, the phase distribution is broad.    

\begin{figure}
\centering
\includegraphics[height=8cm,width=10cm]{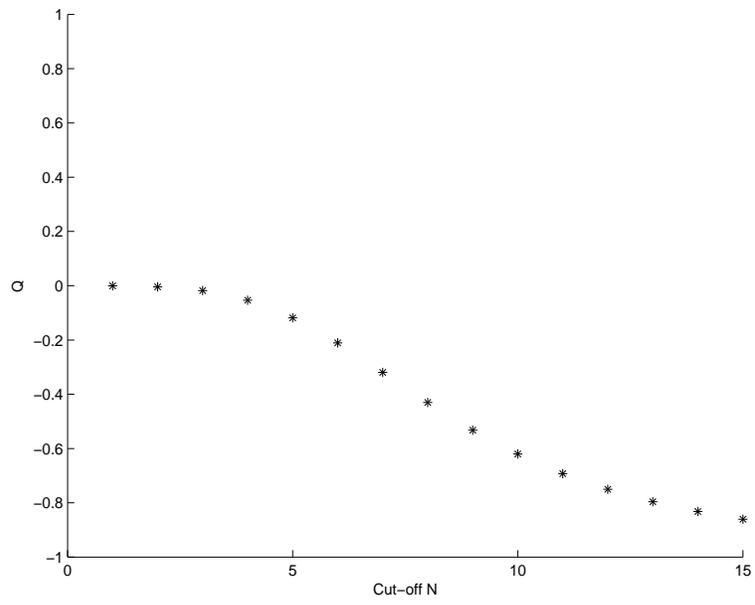}
\caption {$Q$-parameter as a function of the cut-off parameter $N$ for the LTCS of amplitude $\alpha=\sqrt{10}$.}
\label{fig:FigVA}
\end{figure}

The photon-number distribution of the LTCS is not Poissonian.  Whether it is sub-Poissonian or not is inferred from the Q-parameter introduced by Mandel\cite{mandelQ}, 
\beqn
Q=\frac{\la\Delta n\ra^2}{\la\ano\ra}-1=\frac{\vert\alpha\vert^2}{(2+N+\vert\alpha\vert^2)(1+N+\vert\alpha\vert^2)}-1 < 0.
\eeqn 
for the state given in Eq. \ref{bernoulli}, which is the large-$N$ limit of the LTCS.  
This  parameter is used to quantify the deviation from the Poissonian distribution.   
 Consequently, the Q-parameter assumes negative values signalling the sub-Poissonian nature of the LTCS.  Though the above closed form expressions have been obtained taking $N$ to be large,  numerical evaluation of $Q$ for the exact LTCS shows that the states are indeed sub-Poissonian.  The $N$-dependence of the Q parameter is shown in Fig. \ref{fig:FigVA}.  It is seen that for all values of $N$ the Q-parameter is negative.  

\subsection{Entanglement Potential}

     Nonclassicality of a quantum state is amenable to quantification and many  measures of nonclassicality are known: nonclassical distance\cite{hillery, wunsche}, nonclassical depth\cite{ncdepth},  phase-space volume of the negative regions of the Wigner distribution\cite{wigner_neg}, etc.    Of these, the criterion based on the Wigner function is measurable as the Wigner function of a quantum state can be constructed by measuring a  suitably chosen set of observables\cite{risken, meyer, leonhardt}.  Recently, Asboth {\it et al} proposed a measure of nonclassicality of any quantum state of a single-mode of the electromagnetic field\cite{asboth}.  Beam-splitter is a device in which two input modes interact to generate two output modes.  In principle, due to the interaction, the output modes can be entangled.  
A necessary condition for a  beam-splitter to generate entangled output state is that at least one of the inputs should be a nonclassical\cite{kim, wang}.   This feature has been used to define a new measure of nonclasciality called  entanglement potential.   It is the amount of entanglement between the two output modes when the input ports are fed with states $\vert 0\ra\vert\psi\ra$, where $\vert 0\ra$ is the vacuum of one of the modes and the $\vert\psi\ra$ is the state of field input in the other mode. 
Though both the LTCS and UTCS are nonclassical states, it is important to compare the entanglement potentials of the two states.   Each of the input ports of a beam-splitter is modelled as a mode. The unitary operator representing the action of a beam-splitter is $\exp[\Gamma\aco\bao-\Gamma^*\aao\bco]$, $\Gamma$ represents the complex transmittance of the beam splitter\cite{welsch}. Here $\aao$ and $\aco$ are the annihilation and creation operators for one of the input ports. For the other input port, the corresponding operators are $\bao$ and $\bco$.  The operators $K_+=\aco\bao$, $K_-=\aao\bco$ and $K_0=(\ano-\bno)/2$ are generators of the SU(1,1) algebra.  One form of disentangled representation of the unitary operator is
\beqn
\exp(\Gamma K_+-\Gamma^* K_+)=\exp(2\ln\vert\Gamma\vert K_+)\exp(-e^{-i\theta}\tan\vert\Gamma\vert K_0)\exp(e^{i\theta}\tan\vert\Gamma\vert K_-),
\eeqn
where $\Gamma=\vert\Gamma\vert\exp(i\theta)$\cite{disent}.  If vacuum is the input in one of the input ports and $\vert\psi\ra=\sum_{n=0}^\infty c_n\vert n\ra$ in  the other input port, the output state $\vert\chi\ra$ is
\beqn
\vert\chi\ra=\sum_{n=0}^\infty c_n\exp(nB)\sum_{k=0}^n A^k \sqrt{^nC_k}\vert n-k\ra\vert k\ra.
\eeqn 
Then the reduced density operator for the $a$-mode is
\beqn\label{rden}
\rho_{a}=\sum_{n,m=0}^\infty c_n c^*_m\exp(B(n+m))\sum_{k=0}^{\hbox{min}(n,m)}\frac{\vert A\vert^{2k}}{k!}\sqrt{^nC_k}\sqrt{^mC_k}\vert n-k\ra\la m-k\vert,
\eeqn
where $B=-e^{-i\theta}\tan\vert\Gamma\vert$ and $A=2\log\sec\vert\Gamma\vert$.  For 50-50, symmetric  beam splitter, $\Gamma=i\pi/4$.
The coefficients $c_n$ are those of the LTCS or UTCS depending on which state of the field is used in the input port of the beam-splitter.

\begin{figure}
\centering
\includegraphics[height=8cm,width=10cm]{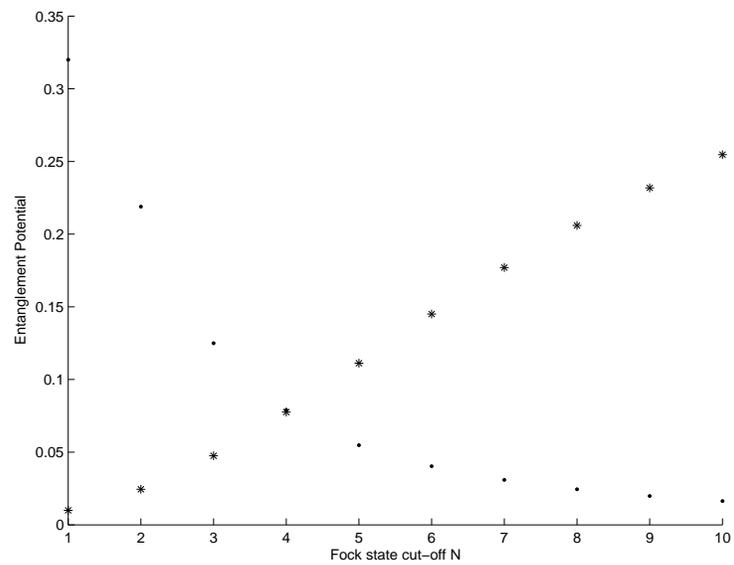}
\caption {Entanglement potential as a function of the cut-off $N$ for the UTCS (dots) and LTCS (star) of amplitude $\alpha=\sqrt{10}$.}
\label{fig:FigVI}
\end{figure}

A measure of entanglement for two-mode, pure states  is the mixedness of the reduced density operator of one of the subsystems, and  linear entropy $1-Tr[\tilde\rho_a^2]$, quantifies the mixedness\cite{bennett}.  Hence, the entanglement between the output field modes is measured by the linear entropy of one of the output modes. The amount of entanglement introduced by the beam-splitter is a measure of nonclassicality of the input field\cite{asboth}.  In Fig. \ref{fig:FigVI}, the linear entropy of the reduced density operator for the output from the beam-splitter is shown.  With the UTCS and vacuum as the input fields, the linear entropy decreases as the number of Fock states in the UTCS increases.  Higher the cutoff, the UTCS approximates the coherent state $\vert\alpha\ra$; consequently, the output fields are less entangled.  The entanglement potential of the LTCS increases as the cutoff $N$ increases, a feature shared by the PACS as well\cite{usha}.    As $N$ increases, the LTCS is essentially the number state $\vert N\ra$, as evident from the large $N$ limit of Eq. \ref{bernoulli} and the entanglement potential of the number state increases with $N$.   Since the nonclassicality of the UTCS decreases with $N$ and that of the LTCS increases, there exists a value of the cutoff parameter at which the states have nearly the same entanglement potential.   If amplitude $\alpha=\sqrt{10}$, the entanglement potential of the two states are nearly equal if the cut-off parameter $N=4$.

\begin{figure}
\centering
\includegraphics[height=8cm,width=10cm]{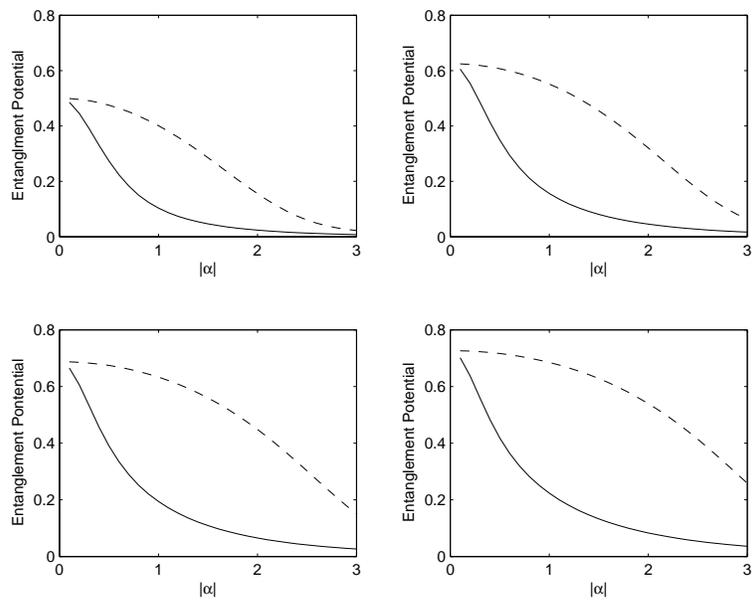}
\caption {Entanglement potential as a function of $\vert\alpha\vert$ for the PACS(continuous) and LTCS (dash).  Figs. (a)-(d) correspond to the lower cutoff values $N$=1,2,3 and 4 respectively. }
\label{fig:FigVII}
\end{figure}

             Recently, the entanglement potential of the PACS has been studied in detail\cite{pacsep}.  Since the LTCS and PACS of equal order are defined on the same sector of the Hilbert space, it is of interest to compare the entanglement potential of these states. The expression for the reduced density  operator given in Eq. \ref{rden} is used with the coefficients $c_n$ being those of the PACS given in Eq. \ref{pacs}.  In Fig. \ref{fig:FigVII}, the entanglement potential of the LTCS of order $N$ is compared with the PACS of same order for $N=$ 1 2,3 and 4.  It is seen that the LTCS has higher entanglement potential compared to the PACS.  This, in turn, means that the LTCS are better suited for generating entangled output modes in a beam-splitter than the PACS.  Currently, a comparison of the two states in the context of quantum  teleportation is being investigated, the results of which will be  presented elsewhere.

\section{Summary}

  The states obtained by removing a set of contiguous, low energy  Fock states from the coherent states of the electromagnetic field are non-Gaussian and nonclassical. These truncated coherent states possess equal uncertainties in the two quadratures, and  do not exhibit squeezing.  Additionally, if the cutoff parameter $N$ is less than $\vert\alpha\vert^2$, the product of the uncertainties attains its minimum, like that of the coherent states.  In this approximation, these states are minimum uncertainty states whose average photon number is large (because $\vert\alpha\vert$ is large), but the photon statistics is sub-Poissonian.  
These states provide a resolution of identity for the subspace spanned by the Fock states $\vert N+1\ra,~\vert N+2\ra,~\vert N+3\ra, \cdots$.  
    The noncalssicality, quantified by the entanglement potential,  of the truncated  states increases as the cutoff $N$ increases.  This is related to the fact that if $N>\vert\alpha\vert^2$, the states are well approximated by a superposition the successive Fock states $\vert N+1\ra$ and $\vert N+2\ra$, which is the generalized Bernoulli state.  Further, the entanglement potential of the truncated coherent states is higher than that of the  photon-added coherent states of same amplitude and order.  The photon-added coherent states and the lower truncated coherent states are the limiting cases of a  suitably deformed photon-added coherent state, wherein the deformation is effected by a function of the number operator.  The photon-added coherent states exhibit squeezing which is absent in truncated coherent states. The states which interpolate between these two liming cases show squeezing. Further, the entanglement potential of truncated coherent states is always higher than that of the photon-added coherent states.  The Jaynes-Cummings hamilotnian without rotating wave approximation is the type of interaction required to produce these interpolating states.

\newpage

\end{document}